\documentclass[twocolumn]{revtex4}

\usepackage{graphicx}
\usepackage{subfigure}
\usepackage{dcolumn}
\usepackage{bm}

\begin{document}

\title{Entanglement and purity of single- and two-photon states}
\author{Jun O.S. Yin and S.J. van Enk}
\affiliation{Oregon Center of Optics, University of Oregon}
\date{\today}

\begin{abstract}
Whereas single- and two-photon wave packets are usually treated as
pure states, in practice they will be mixed. We study how
entanglement created with mixed photon wave packets is degraded.
We find in particular that the entanglement of a delocalized
single-photon state of the electro-magnetic field is determined
simply by its purity. We also discuss entanglement for two-photon
mixed states, as well as the influence of a vacuum component.
\end{abstract}


\maketitle

\section{Introduction}
Consider an entangled state containing one or more photons. By how
much is the entanglement degraded when the photon wave packets are
described by mixed rather than pure states? For example, suppose
Alice and Bob are given a two-photon {\em polarization-entangled}
state \cite{OM,SSR,KWZ,KW,UR}---say the singlet state---but they
are not told what the color of the photons is. All they know is
the photons are either both blue or both green. Does their
ignorance reduce the amount of entanglement they possess? The
answer in this case is negative: the entanglement is still one
ebit, even though the overall state is mixed. After all, they
could, in principle at least, apply a {\em local} measurement on
each photon that measures its color but not its polarization. That
way they are guaranteed to end up with a pure maximally entangled
state. The fact that polarization and color are independent
degrees of freedom is crucial here.

Suppose now Alice and Bob are given a {\em mode-entangled} state
\cite{Z,E3,S} containing merely a single delocalized photon
\cite{TW,H,EXPB,EXPG,EXPH,E5}
\begin{equation}
|\psi\rangle:=\frac{|0\rangle_A|1\rangle_B+|1\rangle_A|0\rangle_B}{\sqrt{2}},\label{psi}
\end{equation}
where $A$ and $B$ denote specific modes in Alice's and Bob's labs
 and $|0\rangle$ and $|1\rangle$ denote Fock states
with zero and one photon, respectively. The notation used implies
that modes $A$ and $B$ are well-defined. But suppose that Alice
and Bob actually do not know what the color of the single
delocalized photon is, green or blue (each with 50\% probability).
Or suppose they do not know the polarization of the photon, only
that it is either left-hand or right-hand circularly polarized.
Then they really should ascribe a mixed state to their field
modes: an equal mixture of $|\psi\rangle$ and the similar state
\begin{equation}
|\psi'\rangle:=\frac{|0\rangle_{A'}|1\rangle_{B'}+|1\rangle_{A'}|0\rangle_{B'}}{\sqrt{2}},
\end{equation}
where the primed modes refer to modes of different color or
different polarization. In Section \ref{single} we will find the
logarithmic negativity \cite{VW,P} of this mixed state to be
$E_{{\cal N}}=\log_2(1+\sqrt{1/2})<1$, so that in this case
Alice's and Bob's state does lose some of its glamour [a pure
state of the form (\ref{psi}) would, obviously, contain one ebit of
entanglement]. Note that, indeed, Alice and Bob cannot use the
same {\em local}  filtering measurement of frequency to filter
their state to a pure entangled state: as soon as a photon is
detected on, say, Bob's side, the state collapses to either
$|0\rangle_A|1\rangle_B$ or $|0\rangle_{A'}|1\rangle_{B'}$, and
not to the desired pure entangled state $|\psi\rangle$ or
$|\psi'\rangle$. Alternatively, a {\em nonlocal} filtering
measurement of color could upgrade Alice's and Bob's state to a
pure entangled state, but that nonlocal operation could and would
increase the amount of entanglement. The distinguishing feature of
this example compared to the previous one is that color or
polarization and mode are dependent degrees of freedom. More
precisely, color and polarization are part of what defines a mode.
(Note, by the way, 
that we can, of course, calculate the logarithmic negativity of the state in the first example as well. The result (see Appendix) is that it indeed equals unity.)
\begin{figure}
\begin{center}
\includegraphics[width=240pt,height=135pt]{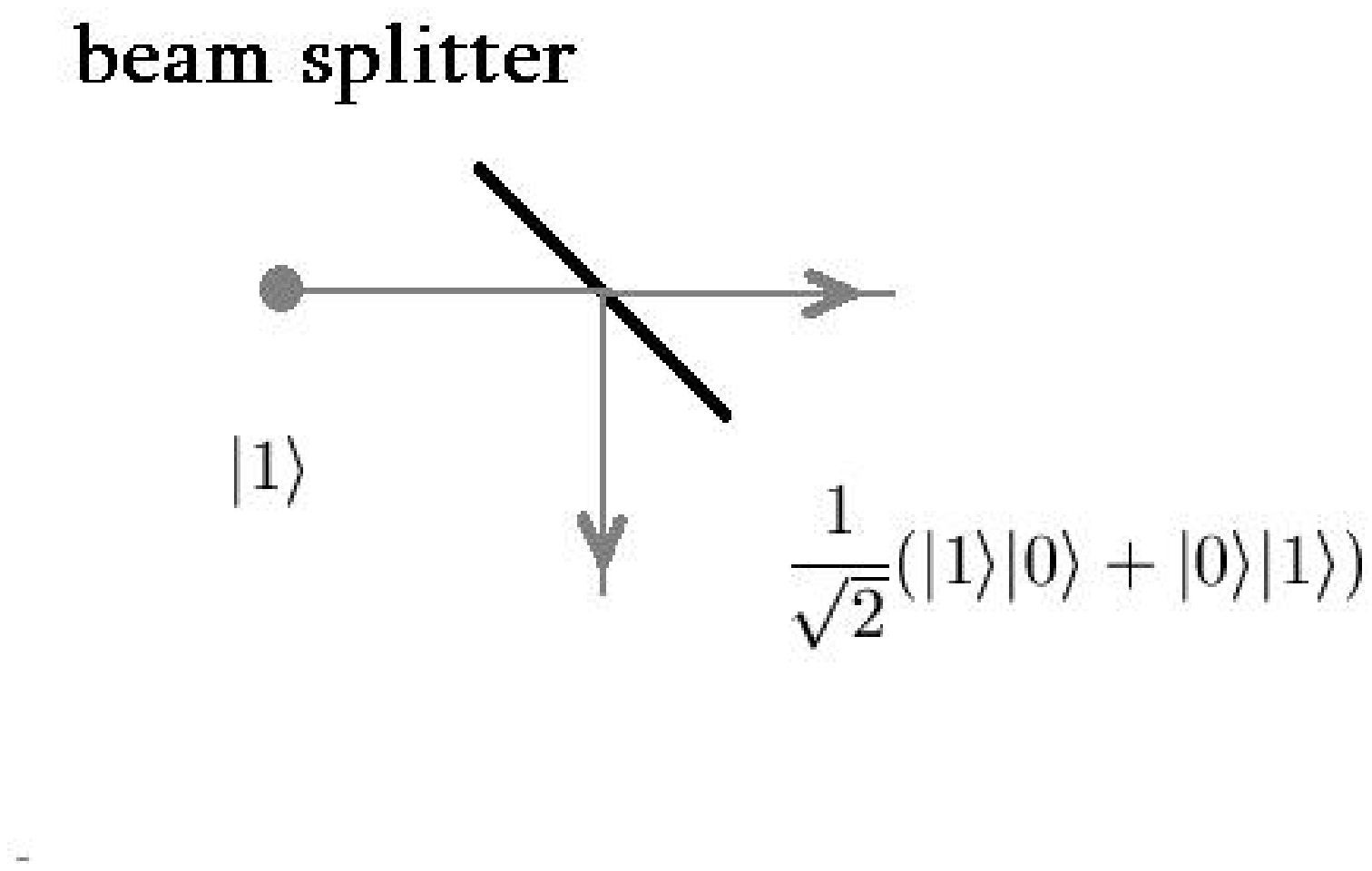}\\
\caption{A single photon impinges on a 50/50 beam splitter and one
ebit of entanglement is created between the two output ports.
How does the entanglement change when the input photon is in a mixed state?\\
 {\bf Answer:} When we use the logarithmic negativity \cite{VW,P}
 as our entanglement monotone of choice we find
 $E_{{\cal N}}=\log_2(1+\sqrt{{\rm purity}}$), in terms of the
 purity Tr$\rho^2$ of the input state (see Section \ref{single}).
}\label{entgen}
\end{center}
\end{figure}

\begin{figure}
\begin{center}
\includegraphics[width=240pt,height=135pt]{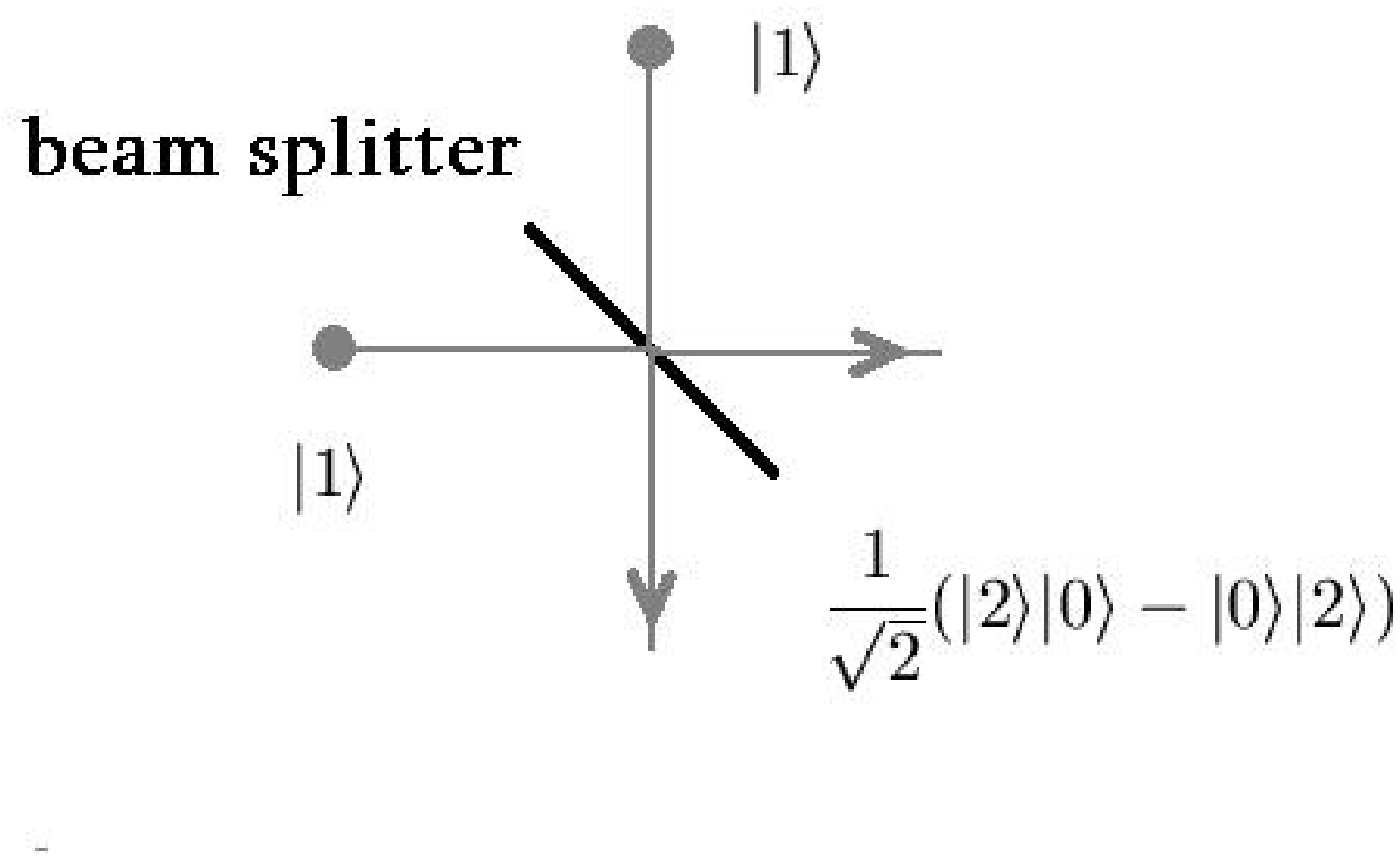}\\
\caption{Two  {\em identical} photons impinge on a 50/50 beam splitter: what is the entanglement at the output? \\
{\bf Answer:}  1 ebit, similarly to the answer in Fig.\ref{entgen}: the output is a delocalized photon {\em pair}.
}\label{entgen1}
\end{center}
\end{figure}

\begin{figure}
\begin{center}
\includegraphics[width=240pt,height=135pt]{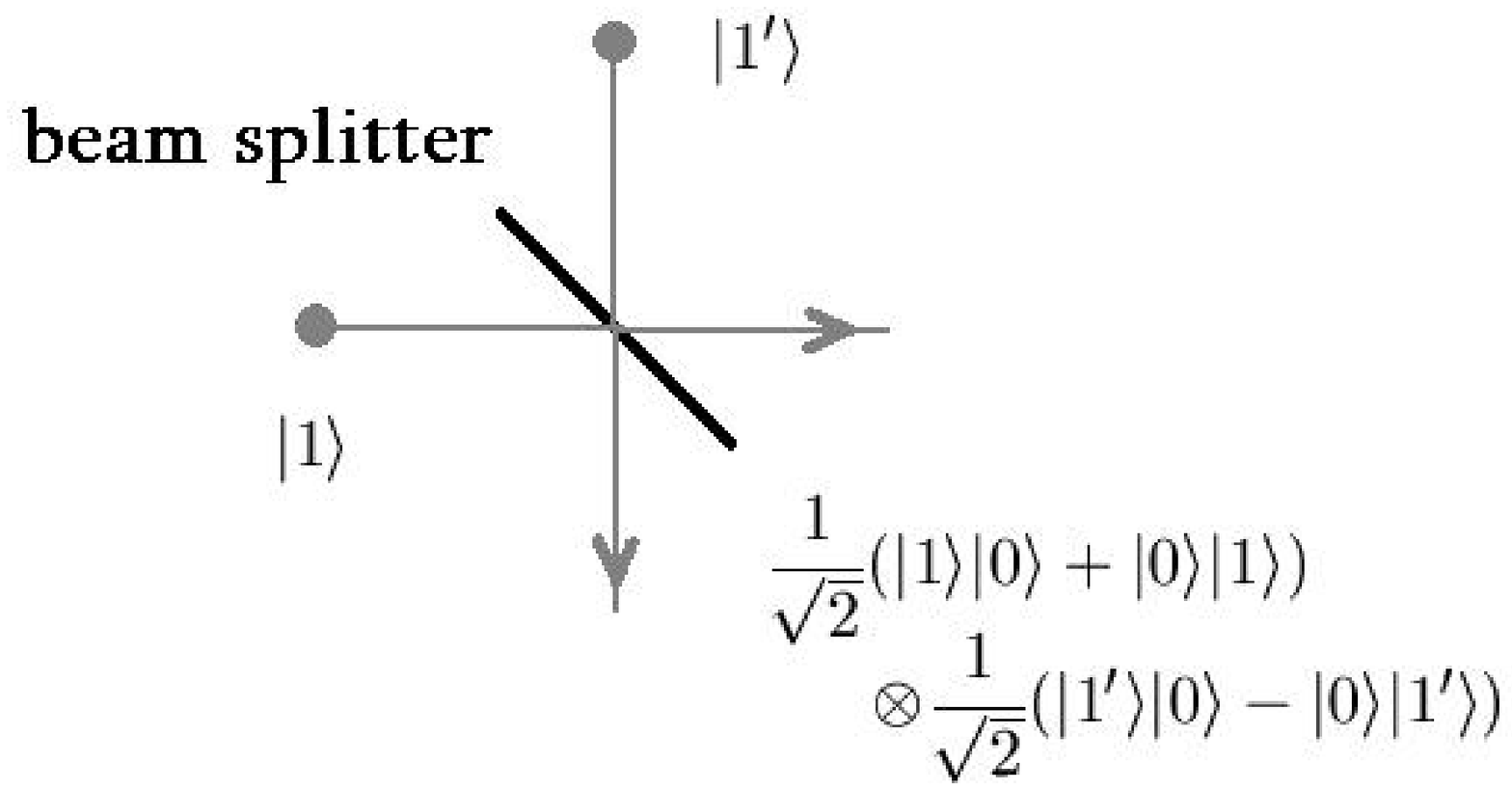}\\
\caption{Two  {\em distinguishable} photons (with, say, orthogonal
polarizations or different colors)  impinge on a 50/50 beam splitter: what is the entanglement at the output? \\
{\bf Answer:} 2 ebits (2 equivalent versions of an entangled delocalized single-photon state) .
}\label{entgen3}
\end{center}
\end{figure}

\begin{figure}
\begin{center}
\includegraphics[width=240pt,height=135pt]{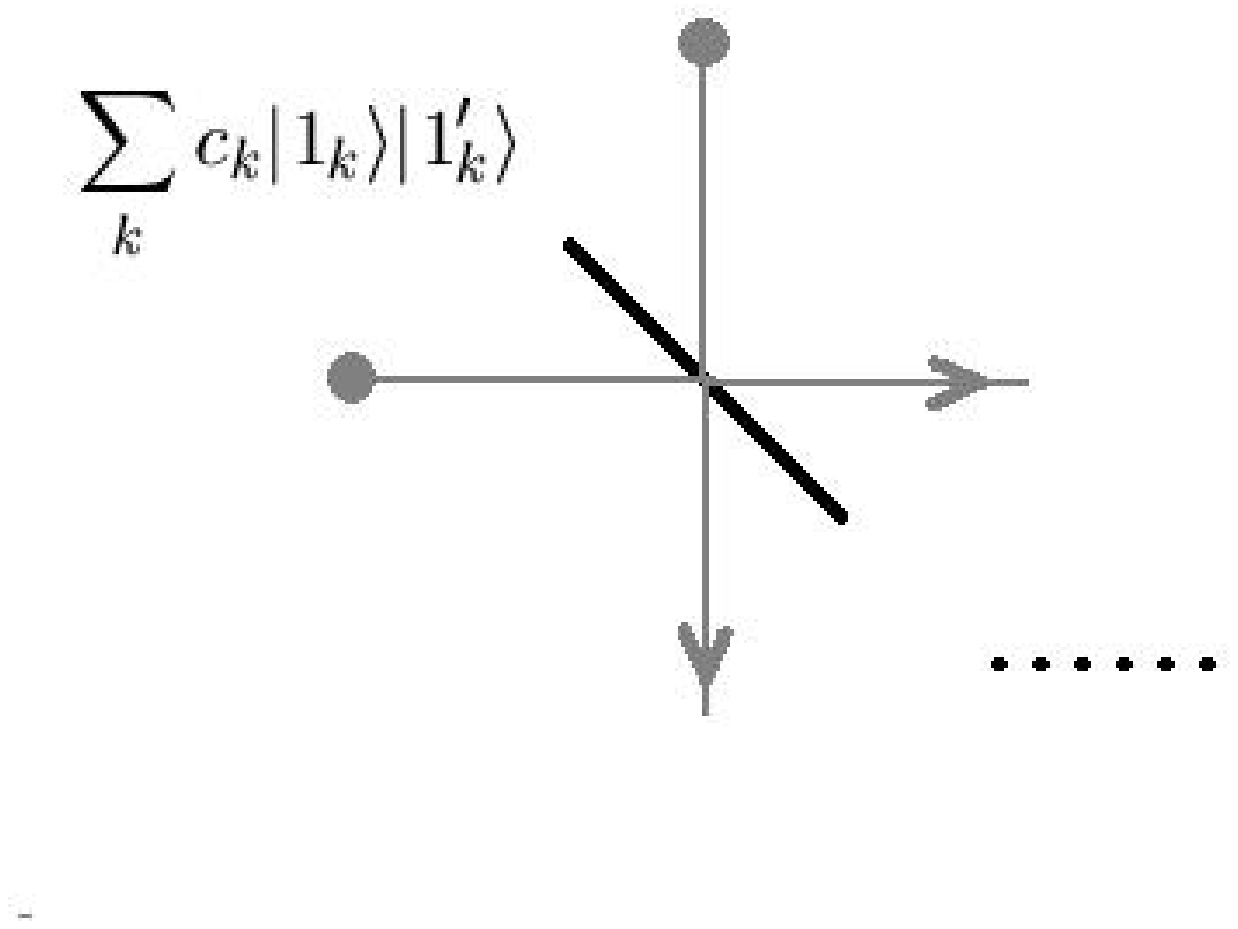}\\
\caption{Two  distinguishable photons with orthogonal
polarizations impinge on a 50/50 beam splitter. When there is
already entanglement between the
polarization degrees of freedom in the input state, what is the entanglement at the output?\\
 {\bf Answer:} $E_{\rm out}=2+E_{\rm in}/2$ for the entropy of entanglement,
 generalizing the answer illustrated in Fig.~\ref{entgen3} (see Section \ref{two}).\\
  How does this answer change when the input state is mixed?\\
  {\bf Answer:} it's complicated \ldots see Section \ref{two} for details.
}\label{entgen2}
\end{center}
\end{figure}
The purpose of this paper is to continue investigating questions
of this sort: by how much is the entanglement of single- or
two-photon states degraded when the photon wave packets are not
pure but mixed? See Figs.~\ref{entgen}--\ref{entgen2} for typical
examples of questions considered in the present paper. The
motivation for this research is, of course, the simple fact that
typically any photon produced in an experiment is represented by a
mixed state \cite{GW,UR,LWKR,RNBW}. For instance, even if  one's
source produces a Fourier-limited wave packet, in practice one
will not know exactly the timing of the wave packet, or the exact
central frequency, or the exact width. For another example,
consider a single photon heralded by detection of the other photon
of a down-converted pair of photons. Whenever there is some
entanglement between the two photons, tracing out one photon
necessarily leaves the remaining photon in a mixed state.
Our study complements those in Refs.~\cite{humble07,humble08}
where the effects of spectral entanglement on polarization degrees of freedom are investigated.

In Section \ref{prelim} we start out by collecting some useful
results about the description of single- and two-photon wave
packets to be used in later sections. In Section \ref{single} we
focus our attention on entangled states that can be generated by
splitting a single photon on a 50/50 beam splitter. Ideally that
leads to an output state with one ebit of entanglement, but, as we
will show, the entanglement resulting from splitting a {\em mixed}
single-photon input state is less than one ebit but turns out to
be a simple function of its purity. In the same Section we will
also consider the more realistic case of a nonzero vacuum
component of the state of the field and its effect on
entanglement. In Section \ref{two} we consider typical mixed
states of two orthogonally polarized photons arising from type-II
down conversion, and in that case too we consider the effects of
the presence of a (large) vacuum component. We also compare the
mode entanglement one obtains by splitting both photons on a 50/50
beam splitter with the entanglement that may already be present in
the input state between two orthogonally polarized modes, and find
one may increase the amount of entanglement that way. We conclude
the paper with a summary that also discusses some possible
extensions of the present
 work.

\section{Single- and two-photon wave packets: preliminaries}\label{prelim}

\subsection{Single photons}

Consider a single photon of a definite polarization propagating in
a well-defined direction. Then a pure state can be described in terms
of continuous modes \cite{BLP} as
\begin{eqnarray}
  |1_\psi\rangle=\int dt\widetilde{\psi}(t)a^\dagger(t)|{\rm
  v}\rangle,\label{p1time}
\end{eqnarray}
where $|{\rm v}\rangle$ is the vacuum state, $a^\dagger(t)$ an
operator that creates a photon at time $t$ and
$\widetilde{\psi}(t)$ the temporal mode function of the wave packet.
Often it is more useful to Fourier transform this representation into
frequency space, and describe the same state by
\begin{eqnarray}
  |1_\psi\rangle=\int d\omega \psi(\omega)a^\dagger(\omega)|{\rm
  v}\rangle.\label{p1freq}
\end{eqnarray}
Typically, the function $\psi(\omega)$ will be appreciable only in
a small bandwidth $\sigma$ around a central frequency $\omega_0$,
with $\sigma\ll\omega_0$, and the wave packet (\ref{p1freq})
describes a quasi-monochromatic photon. Since the creation
operators bear the relation
\begin{eqnarray}
  a^\dagger(t)=\frac{1}{\sqrt{2\pi}}\int d\omega a^\dagger(\omega)
  e^{-i\omega t},
\end{eqnarray}
the spectral shape $\psi(\omega)$ is thus completely determined by
the temporal mode $\widetilde{\psi}(t)$ in the sense that
\begin{eqnarray}
  \psi(\omega)=\frac{1}{\sqrt{2\pi}}\int
  dt\widetilde{\psi}(t)e^{-i\omega t}.
\end{eqnarray}
The single-photon states displayed so far are pure. Mixed states
of single photons arise, e.g., when they are part of a
multipartite system and we trace over the other parties; or if one
of the transverse degrees of freedom of the photon is traced out;
or if we are simply ignorant about one or more of the properties
of the photon. In any case the mixed state of a single photon is
described in terms of the density matrix
\begin{eqnarray}
  \rho_1=\int d\lambda P(\lambda)|1_{\psi_\lambda}\rangle
  \langle1_{\psi_\lambda}|.\label{m1}
\end{eqnarray}
Here $\lambda$ stands for any parameter or combination of
parameters that can possibly be involved in the mode function,
although no concrete form is given yet.  Some good examples of
what $\lambda$ could stand for are arrival time ("time jitter"),
central frequency ("frequency jitter"), the width of the
(Fourier-limited) wave packet, etcetera \cite{LWKR}. When writing the
mixed state in terms of parameters with such a clear physical
meaning, the states $|1_{\psi_\lambda}\rangle$'s will in general
not be orthogonal. However, we can always diagonalize the density
matrix, and rewrite it as a discrete sum involving orthogonal states,
\begin{eqnarray}
  \rho_1=\sum_{k} p_k|1_k\rangle\langle1_k|.\label{single_diag}
\end{eqnarray}
Here $p_k$ and $|1_k\rangle$ are eigensolutions to
\begin{eqnarray}
  \rho_1|1_k\rangle=p_k|1_k\rangle\label{single_diag1}
\end{eqnarray}
with
\begin{eqnarray}
  \langle1_j|1_k\rangle=\delta_{jk},\label{single_diag2}
\end{eqnarray}
and $\sum_k p_k=1$. For our purpose of calculating entanglement of
single-photon states, the latter representation is often more
useful.

\subsection{Two photons}

Now consider states of exactly two photons. Since we have in mind
the two-photon component of the state produced by type-II down
conversion \cite{HM,OZWM,GW}, we assume the photons have
orthogonal polarizations. We will simply indicate the ordinary and
extraordinary polarizations by "H" and "V". Pure and mixed states
for such photon pairs can then be expressed as
\begin{eqnarray}
  |2_\psi\rangle=\int d\omega\int d\omega'\psi(\omega,\omega')
  a_H^\dagger(\omega)a_V^\dagger(\omega')|{\rm v}\rangle,
  \label{p2}
\end{eqnarray}
and
\begin{eqnarray}
  \rho_{2}=\int d\lambda P(\lambda)|2_{\psi_\lambda}\rangle
  \langle2_{\psi_\lambda}|,\label{m2}
\end{eqnarray}
respectively.
The commutators of the mode operators are
\begin{eqnarray}
  [a_k(\omega),a_{k'}^\dagger(\omega')]=\delta_{k,k'}\delta(\omega-\omega'),
\end{eqnarray}
where $\{k,k'\}=\{H,V\}$.

It is well-known that for a pure state of a bipartite system there
is always a discrete Schmidt decomposition. In our case this
allows us to rewrite
\begin{eqnarray}
  |2_\psi\rangle=\sum_k\sqrt{\lambda_k}h_{k}^\dagger
  v_{k}^\dagger|{\rm v}\rangle.\label{Schmidt}
\end{eqnarray}
For a state like Eq.~(\ref{p2}) we can explicitly define the new creation
operators as
\begin{eqnarray}
  h_{k}^\dagger&=&\int d\omega\varphi_k(\omega)a_H^\dagger(\omega),\nonumber\\
  v_{k}^\dagger&=&\int
  d\omega\phi_k(\omega)a_V^\dagger(\omega),\label{hv_operator}
\end{eqnarray}
where $\lambda_k$, $\varphi_k$ and $\phi_k$ can be obtained by
solving the eigenvalue problems \cite{LWE,PBP}
\begin{eqnarray}
  \int d\omega'\widetilde{\rho}_A(\omega,\omega')\varphi_k(\omega')
  =\lambda_k\varphi_k(\omega),\nonumber\\
  \int d\omega'\widetilde{\rho}_B(\omega,\omega')\phi_k(\omega')
  =\lambda_k\phi_k(\omega),
\end{eqnarray}
with the ``reduced density matrices'' given by
\begin{eqnarray}
  \widetilde{\rho}_A(\omega,\omega')=\int
  d\omega''\psi(\omega,\omega'')\psi^\ast(\omega',\omega''),\nonumber\\
  \widetilde{\rho}_B(\omega,\omega')=\int
  d\omega''\psi(\omega'',\omega)\psi^\ast(\omega'',\omega').
\end{eqnarray}
The mode operators $h_{k}^\dagger$'s and $v_{k}^\dagger$'s satisfy
the standard commutation relations
\begin{eqnarray}
  &&[v_{j},h_{k}^\dagger]=0,\nonumber\\
  &&[v_{j},v_{k}^\dagger]=[h_{j},h_{k}^\dagger]=\delta_{jk}.
\end{eqnarray}

\subsection{Mode entanglement}

In the rest of the paper we will calculate mode entanglement
between field modes \cite{Z,E3,S} that are spatially separated
\cite{Sp}. For that purpose we expand the density matrix in an
appropriate orthonormal basis of Fock states, including states
with no photons, single-photon states $|1_k\rangle$ as defined
above, etc.

We use two standard measures of entanglement in this paper: one
useful entanglement measure, but only valid for pure bipartite
states, is the entropy of entanglement \cite{BDSW}. For example the
mode entanglement between modes of orthogonal polarization of the
pure state $|2_{\psi}\rangle$ of Eq.~(\ref{p2}) is defined in
terms of the Schmidt coefficients appearing in Eq.~(\ref{Schmidt})
as
\begin{eqnarray}
  E(|2_\psi\rangle)=-\sum_k{\lambda_k\log_2\lambda_k}.
\end{eqnarray}
The other entanglement monotone we will make use of is the
logarithmic negativity \cite{VW,P}, defined in terms of the partial
transpose (PT) of a matrix.  For example, for a two-photon system
PT is easily defined if we expand its density matrix in the basis
spanned by $|k\rangle_1|l\rangle_{21}\langle m|_2\langle n|$, with
$|k\rangle_1$, $|l\rangle_2$, $|m\rangle_1$ and $|n\rangle_2$
referring to orthogonalized single-photon states respectively. Then
we have
\begin{eqnarray}
  \rho&\stackrel{PT}{\rightarrow}&\rho^{\Gamma},\\
  \rho_{klmn}^{\Gamma}&=&\rho_{knml},
\end{eqnarray}
$\rho_{klmn}$ being the matrix element. We then solve for the
eigenvalues of $\rho^{\Gamma}$, ending up with a series of real
$e_k$'s, since both $\rho$ and $\rho^{\Gamma}$ are Hermitian. The
logarithmic negativity \cite{VW,P} is defined as
\begin{eqnarray}
  E_{{\cal N}}(\rho)&=&\log_2{\|\rho^{\Gamma}\|_1}=\log_2{\sum_k{|e_k|}}.
\end{eqnarray}
It is worthwhile to notice that the absolute sum of all
eigenvalues $\sum_k{|e_k|}$ equals 1 plus twice the absolute value
of the sum of all negative eigenvalues of $\rho^{\Gamma}$.

\section{Mode entanglement of single-photon states}\label{single}

\subsection{No vacuum component}
\subsubsection{General results}
When an optical field containing exactly one photon in a {\em pure} state is split on a 50/50 beam splitter, the output state looks like
\begin{eqnarray}
  \frac{1}{\sqrt{2}}(|0\rangle|1\rangle+|1\rangle|0\rangle)
\end{eqnarray}
and possesses exactly one ebit of entanglement
\cite{EXPH,EXPG,E5}. If, on the other hand, the photon entering one input port of the
beam splitter is {\em mixed}, the calculation of entanglement in the
output state is a little more complicated, but the problem can
still be solved analytically. We consider single polarization only
and start by an input state that is already expanded in its
diagonal form, Eq.~(\ref{single_diag})
\begin{eqnarray}
  \rho_{\rm in}=\sum_{k=1}^n p_k|1_k\rangle\langle1_k|.
\end{eqnarray}
We put it on a 50/50 beam splitter, which in the Heisenberg picture
transforms the mode operators for the input ports
$a^\dagger,b^\dagger$ into those for the output
$c^\dagger,d^\dagger$ in the following way
\begin{eqnarray}
  \left(%
  \begin{array}{c}
    a^\dagger \\
    b^\dagger \\
  \end{array}%
  \right)=\frac{1}{\sqrt{2}}
  \left(%
  \begin{array}{cc}
    1 & 1 \\
    1 & -1 \\
  \end{array}%
  \right)
  \left(%
  \begin{array}{c}
    c^\dagger \\
    d^\dagger \\
  \end{array}%
  \right).
\end{eqnarray}
In this picture the state stays unchanged, but when expressed
in terms of creation operators at the output ports it looks
different:
\begin{eqnarray}
  \rho_{\rm out}&=&\rho_{\rm in}\nonumber\\
  &=&\sum_{k=1}^n p_k|1_k\rangle_a\otimes_a\langle1_k|\nonumber\\
  &=&\frac{1}{2}\sum_{k=1}^n p_k(|1_k\rangle_c|0\rangle_d
  +|0\rangle_c|1_k\rangle_d)\nonumber\\
  &&\otimes(_c\langle1_k|_d\langle0|+_c\langle0|_d\langle1_k|)\\
  &=&\sum_{k=1}^n\rho_k,
\end{eqnarray}
where each submatrix $\rho_k$ can be written out as
\begin{eqnarray}
  \rho_k=\left(%
\begin{array}{cccc}
  0 & 0 & 0 & 0 \\
  0 & p_k/2 & p_k/2 & 0 \\
  0 & p_k/2 & p_k/2 & 0 \\
  0 & 0 & 0 & 0 \\
\end{array}%
\right).
\end{eqnarray}
Here rows and columns correspond to states
$|0\rangle_c|0\rangle_d$, $|0\rangle_c|1_k\rangle_d$,
$|1_k\rangle_c|0\rangle_d$, $|1_k\rangle_c|1_k\rangle_d$ and their
conjugates respectively. Naively, taking the PT simply gives
\begin{eqnarray}
  \rho_k^{\Gamma}=\left(%
\begin{array}{cccc}
  0 & 0 & 0 & p_k/2 \\
  0 & p_k/2 & 0 & 0 \\
  0 & 0 & p_k/2 & 0 \\
  p_k/2 & 0 & 0 & 0 \\
\end{array}%
\right).
\end{eqnarray}
 In the total density matrix context,
the two $p_k/2$ on the diagonal remain independent of other
$\rho_k$'s. These give rise to nonnegative eigenvalues of
$\rho^{\Gamma}$. Care must be taken, however, with the two
off-diagonal elements. Those matrix elements share the vacuum
state $|0\rangle_c|0\rangle_d$ with all other submatrices, and
therefore beg an eigenvalue solution to the matrix
\begin{eqnarray}
  \left(%
\begin{array}{cccc}
  0 & p_1/2 & ... & p_n/2 \\
  p_1/2 & 0 & ... & 0 \\
  ... & ... &  & ... \\
  p_n/2 & 0 & ... & 0 \\
\end{array}%
\right).
\end{eqnarray}
It can be shown, with a little effort, that the only two nonzero
eigenvalues of this matrix are
\begin{eqnarray}
  \lambda_{1,2}=\pm\frac{1}{2}\sqrt{\sum_{k=1}^np_k^2}.
\end{eqnarray}
Notice from Eq.~(\ref{single_diag}) that
\begin{eqnarray}
  \sum_{k=1}^n p_k^2={\rm Tr}\rho_{\rm in}^2,
\end{eqnarray}
and that ${\rm Tr}\rho_{\rm in}^2$ is an invariant quantity under
basis transformations, or equivalently, unitary operations.
$\lambda_{1,2}$ can thus be evaluated as
\begin{eqnarray}
  \lambda_{1,2}=\pm\frac{1}{2}({\rm Tr}\rho_{\rm in}^2)^{1/2}.
\end{eqnarray}
Since we learn from the definition that $\|\rho^{\Gamma}\|_1$ equals
half of the absolute value of the sum of all negative eigenvalues, we immediately get
\begin{eqnarray}
  E_{{\cal N}}(\rho_{\rm out})&=&\log_2[1+({\rm
  Tr}\rho_{\rm in}^2)^{1/2}],\label{LNm1}
\end{eqnarray}
as announced in Fig.~\ref{entgen}. (Of course, the purity of the
output state equals that of the input state, as we assumed the
beam splitter to be describable by a unitary operation.)
\subsubsection{Example}

With this conclusion, we attempt to calculate the logarithmic
negativity for  single-photon mixed states for which all wave
packet functions $\psi(\omega)$ are Gaussian-shaped
\begin{eqnarray}
  \psi(\omega)\propto\exp\left(-\frac{(\omega-\omega_0)^2}{\sigma^2}\right),
  \label{wavpack}
\end{eqnarray}
and which is mixed with respect to the arrival time $\tau$ of
wave-packet peaks. Assuming a Gaussian distribution for arrival
times as well, we write
\begin{eqnarray}
  \psi(\omega)&\rightarrow&\psi_\tau(\omega)=\psi(\omega)e^{-i\omega\tau},
  \label{1p_delay}\\
  P(\tau)&\propto&\exp\left(-\frac{\tau^2}{\sigma_\tau^2}\right).\label{mixdist}
\end{eqnarray}
So, explicitly the density matrix is
\begin{eqnarray}
  \rho_{\rm in}&=&A\int d\tau\exp\left(-\frac{\tau^2}{\sigma_\tau^2}\right)\int
  d\omega\int d\omega'\nonumber\\
  &&\exp\left[-\frac{(\omega-\omega_o)^2}{\sigma^2}-\frac{(\omega'-\omega_o)^2}{\sigma^2}+i(\omega'-\omega_0)\tau\right]\nonumber\\
  &&\times a^\dagger(\omega)|{\rm
  v}\rangle\langle{\rm v}|a(\omega'),\label{rhoin1}
\end{eqnarray}
with a normalization coefficient $A$. Since the state is profiled in
spectral space, we have
\begin{eqnarray}
  {\rm Tr}\rho_{\rm in}^2=\int d\omega\langle{\rm
  v}|a(\omega)\rho_{\rm in}^2a^{\dagger}(\omega)|{\rm v}\rangle.
\end{eqnarray}
By using the commutation relation
\begin{eqnarray}
  [a(\omega),a^\dagger(\omega')]=\delta(\omega-\omega')
\end{eqnarray}
and basic algebra we can show that
\begin{eqnarray}
  {\rm Tr}\rho_{\rm in}^2=(1+4\sigma^2\sigma_\tau^2)^{-1/2}.
\end{eqnarray}
The logarithmic negativity after the beam splitter is then according to (\ref{LNm1})
\begin{eqnarray}
  E_{{\cal N}}(\rho_{\rm out})=\log_2[1+(1+4\sigma^2\sigma_\tau^2)^{-1/4}],
\end{eqnarray}
which in either limit $\sigma_\tau=0$ or $\sigma=0$ reduces to
unity. The first limit is just a pure state, which is
easily conceived, while the latter fact is a bit harder to reveal,
but think of $\psi(\omega)$ as $\delta(\omega-\omega_o)$ when
$\sigma\rightarrow0$. Then, according to Eq.~(\ref{rhoin1}) but with a different normalization constant $A'$, we have
\begin{eqnarray}
  \rho_{\rm in}&=&A'\int d\tau\exp\left(-\frac{\tau^2}{\sigma_\tau^2}\right)
  \int d\omega\delta(\omega-\omega_o)e^{-i\omega\tau}\nonumber\\
  &&\int d\omega'\delta(\omega'-\omega_o)e^{i\omega'\tau}
  a^\dagger(\omega)|{\rm v}\rangle\langle{\rm
  v}|a(\omega')\nonumber\\
  &=&A'\int d\tau\exp\left(-\frac{\tau^2}{\sigma_\tau^2}\right)
  a^\dagger(\omega_o)|{\rm v}\rangle\langle{\rm v}|a(\omega_o),
\end{eqnarray}
which is equivalently a pure state. Physically,
$\delta(\omega-\omega_o)$ corresponds to monochromatic light whose
wave function extends homogeneously along the time axis to both
infinities, so that the concept of wave-packet arrival time no
longer applies.

In conclusion, the entanglement of the output state depends only
on the ratio of the ``incoherent'' width of the mixture in time,
$\sigma_\tau$ to the ``coherent'' width in time of each wave
packet $1/\sigma$. For a large incoherent width the entanglement
reduces to zero, as expected.

\subsection{Adding the vacuum}

Real experiments involving single photons typically are described
by a state involving a vacuum component in addition to the
single-photon component. The phase between the vacuum and a
particular single-photon Fock state may or may not be known or
controlled. We may write a state containing the two
Fock states as
\begin{eqnarray}
  \rho_{\rm vac1_{in}}=\int d\varphi f(\varphi)(\sqrt{1-p}|{\rm
  v}\rangle+\sqrt{p}e^{i\varphi}|1\rangle)\nonumber\\
  \otimes(\sqrt{1-p}\langle{\rm
  v}|+\sqrt{p}e^{-i\varphi}\langle1|),
\end{eqnarray}
with $p$ the fixed {\em a priori} probability for the state to contain exactly one
photon. We consider two extreme cases here:
\begin{itemize}
  \item $f(\varphi)\sim\delta(\varphi)$ when the state is pure and
  we simply represent it, instead of $\rho$, as a pure state
  \begin{eqnarray}
    |{\rm vac1_{in}}\rangle=\sqrt{1-p}|{\rm
    v}\rangle+\sqrt{p}|1\rangle;\label{vac_p1_in}
  \end{eqnarray}
  \item $f(\varphi)$ is a flat distribution so that the cross
  terms vanish after the integration. The state is thus reduced to
  \begin{eqnarray}
    \rho_{\rm vac1_{in}}=(1-p)|{\rm v}\rangle\langle{\rm
    v}|+p|1\rangle\langle 1|.\label{vac_m1_in}
  \end{eqnarray}
\end{itemize}
We treat these two cases one by one. The output state of
Eq.~(\ref{vac_p1_in}) after a 50/50 beam splitter is
\begin{eqnarray}
  |{\rm vac1_{out}}\rangle=\sqrt{1-p}|00\rangle+\sqrt{\frac{p}{2}}
  (|10\rangle+|01\rangle)\label{vac_p1_out}.
\end{eqnarray}
The entropy of entanglement and the logarithmic negativity are
straightforwardly calculated and the result is
\begin{eqnarray}
  &&E(|{\rm vac1_{out}}\rangle)\nonumber\\
  &=&1-\frac{1}{2}[(1+\sqrt{1-p^2})\log_2(1+\sqrt{1-p^2})\nonumber\\
  &&+(1-\sqrt{1-p^2})\log_2(1-\sqrt{1-p^2})],\label{vac_p1_Eout}
\end{eqnarray}
\begin{eqnarray}
  &&E_{{\cal N}}(|{\rm vac1_{out}}\rangle)=\log_2(1+p).\label{vac_p1_LNout}
\end{eqnarray}
The latter expression is particularly simple. Of course, both
measures of entanglement vary between 0 and 1 for $p$ varying
between 0 and 1. It can be shown that $E_{{\cal N}}$ is always
larger that $E$ when $p\in[0,1]$.

For the mixture, we are at liberty to assume the diagonal
expansion Eq.~(\ref{single_diag}) along with
Eq.~(\ref{single_diag1}) and Eq.~(\ref{single_diag2}). Hence we
have
\begin{eqnarray}
  \rho_{\rm vac1_{out}}&&=(1-p)|00\rangle\langle00|\nonumber\\
  &&+\frac{p}{2}\sum_k\lambda_k
  (|01_k\rangle+|1_k0\rangle)(\langle01_k|+\langle 1_k0|).\label{vac_m1_out}
\end{eqnarray}
The quantities we are interested in are now
\begin{eqnarray}
  Pur(\rho_{\rm
  vac1_{out}})&=&(1-p)^2+p^2\sum_k\lambda_k^2,\label{vac_m1_PURout}\\
  E_{{\cal N}}(\rho_{\rm
  vac1_{out}})&=&\log_2(p+Pur^{1/2}),\label{vac_m1_LNout}
\end{eqnarray}
where $Pur$ denotes the purity of the input state. We see that
Eq.~(\ref{vac_m1_LNout}) generalizes the expression
Eq.~(\ref{vac_p1_LNout}) which is only a special case when
$Pur=1$. Of course, it also generalizes Eq.~(\ref{LNm1}). In
conclusion, $E_{{\cal N}}$ depends only on $p$ and the purity,
which in turn is also affected by the value of $p$.

\section{Entanglement of two-photon states}\label{two}

\subsection{No vacuum component}

A broadband-pumped down-conversion process produces a state whose
two-photon component can be written in the form (\ref{p2}), with
the mode function in the ideal (pure-state) case described as
\cite{GW}
\begin{eqnarray}
  \psi(\omega_o,\omega_e)=\alpha(\omega_o+\omega_e)
  \Phi(\omega_o,\omega_e).\label{modefuncp2}
\end{eqnarray}
Here $\omega_{o,e}$ are the frequencies of the two photons
with ordinary and extraordinary polarization, respectively.
 $\alpha(\omega_o+\omega_e)$ is the pump spectrum envelope, and
the phase-matching function $\Phi(\omega_o,\omega_e)$, after a
great deal of simplification, is
\begin{eqnarray}
  &&\Phi(\bar{\omega}_o+\nu_o,\bar{\omega}_e+\nu_e)\nonumber\\
  &=&{\rm sinc}\{[\nu_o(k'_o-k'_p)+\nu_e(k'_e-k'_p)]L\},\label{sinc}
\end{eqnarray}
where $\nu_{o,e}$ are deviations from the perfect match
frequencies $\bar{\omega}_{o,e}$, and where $L$ is the length of
the nonlinear medium. Moreover, $k'_{p,o,e}$ are the first
derivatives of wave vectors with respect to frequency for pump
photon and outcoming $o$- and $e$- photons respectively at the
perfect phase-matching condition
\begin{eqnarray}
  \omega_p=\bar{\omega}_o+\bar{\omega}_e.
\end{eqnarray}
We still restrict ourselves to Gaussian wave packets only, and so we
assume
\begin{eqnarray}
  \alpha(\nu_o+\nu_e)\sim\exp\left(-\frac{(\nu_o+\nu_e)^2}{\sigma^2}\right),\label{Gauss}
\end{eqnarray}
so that $\psi(\omega_o,\omega_e)$ is real everywhere. In general,
due to our ignorance about the precise timing of the pump pulse or
about its precise central frequency, the state generated will
actually be a mixed state.

We are interested in the entanglement of such a (pure or mixed)
state between the two orthogonally polarized modes. Moreover, just
like in the preceding section, we wish to calculate the
entanglement that results from splitting such a two-photon state
on a 50/50 beam splitter. The resulting entanglement after the
beam splitter is of a different sort, it's entanglement between
the two output modes of the beam splitter, not between orthogonal
polarizations. An interesting question is whether that
entanglement is larger or smaller than the initial polarization
entanglement.

Although we will have to resort to numerical methods to calculate
both types of entanglement, we can analytically determine the
relation between pure-state entanglement before the beam splitter
and that after the beam splitter: Assume that the Schmidt
decomposition of a state [those coefficients can be obtained, in
some approximation, analytically \cite{LW}, but we won't need them
explicitly] described by Eq.~(\ref{modefuncp2}) is
\begin{eqnarray}
  |2_{\rm in}\rangle=\sum_k\sqrt{\lambda_k}h_{k}^\dagger
  v_{k}^\dagger|{\rm v}\rangle,\label{Schdec}
\end{eqnarray}
(Here we have associated $o$-photons with horizontal polarization
and $e$-photons with vertical polarization.) The entropy of
entanglement and logarithmic negativity are thus expressed in
terms of Schmidt coefficients as\cite{VW}
\begin{eqnarray}
  E(|2_{\rm in}\rangle)=-\sum_k\lambda_k\log_2\lambda_k,\\
  E_{{\cal N}}(|2_{\rm in}\rangle)=2\log_2(\sum_k\sqrt{\lambda_k}).
\end{eqnarray}
It can be shown that the state after the beam splitter can be
Schmidt-decomposed similarly as
\begin{eqnarray}
  &&|2_{\rm out}\rangle=\frac{1}{2}|\tilde{2}\rangle_c|0\rangle_d
  -\frac{1}{2}|0\rangle_c|\tilde{2}\rangle_d\nonumber\\
  &&-\sum_k\frac{\sqrt{\lambda_k}}{2}h_{ck}^{\dagger}v_{dk}^{\dagger}|{\rm
  v}\rangle+\sum_k\frac{\sqrt{\lambda_k}}{2}v_{ck}^{\dagger}h_{dk}^{\dagger}|{\rm
  v}\rangle,
\end{eqnarray}
where $h_{ck}^\dagger$, $v_{ck}^\dagger$, $h_{dk}^\dagger$ and
$v_{dk}^\dagger$ are associated with $c_H^\dagger(\omega)$,
$c_V^\dagger(\omega)$, $d_H^\dagger(\omega)$ and
$d_V^\dagger(\omega)$ according to Eqs.\ref{hv_operator}.
$|\tilde{2}\rangle_{c,d}$ stands for a specific two photon state,
take $c$ e.g.,
\begin{eqnarray}
  |\tilde{2}\rangle_c=\sum_k\sqrt{\lambda_k}h_{ck}^\dagger
  v_{ck}^\dagger|{\rm v}\rangle\label{tilde2_disc}
\end{eqnarray}
which in frequency space is actually
\begin{eqnarray}
  |\tilde{2}\rangle_c=\int d\nu_o\int d\nu_e\alpha(\nu_o+\nu_e)
  \Phi(\nu_o,\nu_e)\nonumber\\
  c_H^\dagger(\nu_o)c_V^\dagger(\nu_e)|{\rm
  v}\rangle.\label{tilde2_cont}
\end{eqnarray}
The notation for the state is just a shorthand notation
emphasizing its orthogonality with respect to any single-photon
state or vacuum state. Hence we may conclude
\begin{eqnarray}
  E(|2_{\rm out}\rangle)&=&2+\frac{1}{2}E(|2_{\rm
  in}\rangle),\label{Ea&bBS2p}\\
  E_{{\cal N}}(|2_{\rm out}\rangle)&=&2\log_2(1+2^{E_{{\cal N}}(|2_{\rm
  in}\rangle)/2}).\label{LNa&bBS2p}
\end{eqnarray}
One obvious question is now whether these same relations will
still hold for mixed input and output states.



Since analytical solutions to the two-photon problem seem
impossible without further approximations, even for the pure-state
case (but for an exception see Ref.~\cite{LW}), we therefore turn
to numerical methods where we introduce some standard
approximations that were used before in \cite{LWE,PBP}. Integrals
over continuous frequency are converted into sums over discrete
frequencies, and infinity as the integral limit is replaced by an
artificial cutoff according to
\begin{eqnarray}
  \int_{-\infty}^\infty d\omega\int_{-\infty}^\infty
  d\omega'\rightarrow\Delta\omega\Delta\omega'\sum_{j=1}^n\sum_{k=1}^{n'}.
\end{eqnarray}
For convenience we hence choose $\Delta\omega=\Delta\omega'$ and
$n=n'$. The choice of $\Delta\omega$ is determined by requiring
the integrals to converge numerically. Moreover, we have to choose
a scale for the many frequencies that occur in this problem. Quite
arbitrarily, we have chosen to rescale all quantities with a
dimension of frequency to $\Omega$, defined through the relation
\[
(k'_o-k'_p)L\Omega=2.25.
\]
We also use
\[(k'_e-k'_p)L\Omega=0.63.\]
 These two relations are similar to those used in Ref.~\cite{LWE}.

\begin{figure}
\begin{center}
 \includegraphics[width=240pt,height=135pt]
    {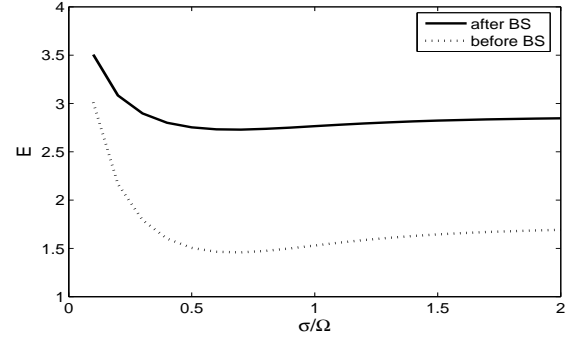}
      \caption{Entropy of entanglement of pure two-photon states
      described by Eqs.~(\ref{modefuncp2})--(\ref{Gauss})
  before and after a 50/50 beam splitter as a function of the dimensionless
  pump width $\sigma/\Omega$.}\label{purEEabBS}
\end{center}
\end{figure}

\begin{figure}
\begin{center}
\includegraphics[width=240pt,height=135pt]{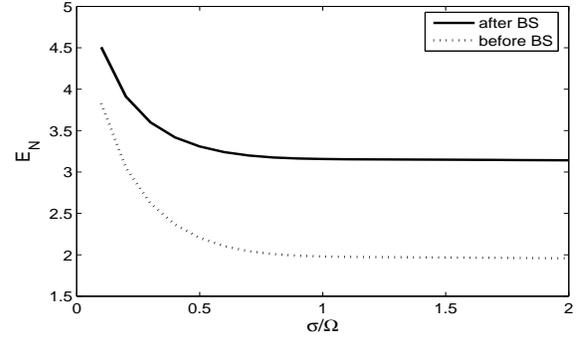}
  \caption{Logarithmic negativity of  pure two-photon states
  described by Eqs.~(\ref{modefuncp2})--(\ref{Gauss})
  before and after a 50/50 beam splitter as a function of the
  dimensionless pump width $\sigma/\Omega$. }\label{purLNabBS}
\end{center}
\end{figure}

In Figs.~\ref{purEEabBS} and \ref{purLNabBS} we plot the
numerically evaluated entanglement for pure states, as a function
of the width $\sigma$ of the Gaussian pump pulse in units of
$\Omega$. We verified the validity of Eq.~(\ref{Ea&bBS2p}) and
Eq.~(\ref{LNa&bBS2p}). Both $\nu_o$ and $\nu_e$ (deviation from
the perfect-match frequencies) are cut off from $-2\Omega$ to
$2\Omega$. This is not driven by questions of numerical
convergence, but by the freedom one has to consider only those
photons in a certain frequency interval. For instance, if one uses
narrow-band detectors then only the entanglement between photons
of frequency within that bandwidth will be relevant. We simply
used cut-off values such that the central peak and two side peaks
of the sinc function (\ref{sinc}) are taken into account.



Next we consider mixed two-photon states in a way that is similar
to what we did for single photon states. We assume Gaussian wave
packets with an uncertainty in arrival time. That is, we introduce
a time-displaced mode function for two frequencies by
\begin{eqnarray}
  \psi(\omega,\omega')\rightarrow\psi_\tau(\omega,\omega')=
  \psi(\omega,\omega')e^{-i(\omega+\omega')\tau},\label{2p_delay}
\end{eqnarray}
and we choose $P(\lambda)$ in Eq.~(\ref{m2}) to take exactly the
same form as Eq.~(\ref{mixdist}) only that $\lambda$ is replaced
by $\tau$. Since the Gaussian shape for $\tau$ falls off fairly
quickly, we decided to extend the integral in our numerical
calculations over the interval $(-2\sigma_\tau,2\sigma_\tau)$. To
be more explicit, after the 50/50 beam splitter the mixed state
looks like
\begin{eqnarray}
  \rho_{2{\rm out}}&=&\int d\tau P(\tau)|2_{\rm out}(\tau)\rangle\langle2_{\rm
  out}(\tau)|,\label{mix2pout}
\end{eqnarray}
where $P(\tau)\sim\exp(-\tau^2/\sigma^2_\tau)$ and
\begin{eqnarray}
  &&|2_{\rm out}(\tau)\rangle=\frac{1}{2}\int d\nu_o\int
  d\nu_e\alpha_\tau(\nu_o+\nu_e)\Phi(\nu_o,\nu_e)\nonumber\\
  &&\times
  e^{-i(\nu_o+\nu_e)\tau}(c_H^\dagger(\nu_o)c_V^\dagger(\nu_e)
  +c_V^\dagger(\nu_e)d_H^\dagger(\nu_o)\nonumber\\
  &&-c_H^\dagger(\nu_o)d_V^\dagger(\nu_e)
  -d_H^\dagger(\nu_o)d_V^\dagger(\nu_e))|{\rm v}\rangle.
  \label{2pout}
\end{eqnarray}
By realizing that the two-photon states can be expanded in their
own space $|2(\tau)\rangle_{c,d}$ instead of single-photon
combination space, the size of the density matrix is greatly
reduced from $(N+1)^4\times (N+1)^4$ to
$(2N+T+1)^2\times(2N+T+1)^2$, $N$ and $T$ being the number of
discretization intervals along $\omega$ and $\tau$ axes
respectively.
\begin{figure}
\begin{center}
  \includegraphics[width=240pt,height=135pt]{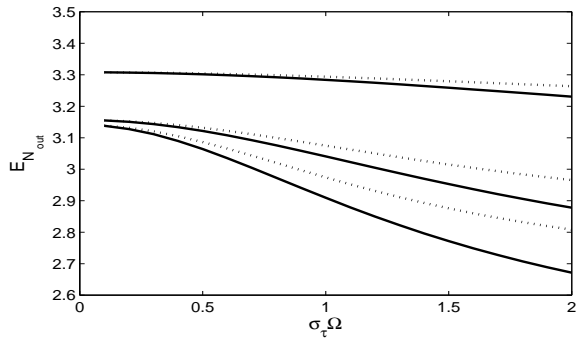}\\
  \caption{Logarithmic negativity of mixed two-photon states (\ref{mix2pout})
  --(\ref{2pout}) after a 50/50 beam splitter vs. the
  dimensionless  mixed-state width $\sigma_\tau\Omega$. The dotted
  curves are calculated by plugging the input state's $E_{{\cal N}}$ into
  Eq.~(\ref{LNa&bBS2p}) and the solid curves are direct numerical
  results. Each set of dotted and solid curves has different
  wave packet width $\sigma$. From top to bottom:
  $\sigma/\Omega=0.5,1,2$.}\label{LN-sigma}
\end{center}
\end{figure}

Fig.~\ref{LN-sigma} shows our numerical results regarding mixed
two-photon states. We display two kinds of curves for three sets
of parameters. The solid curves give the numerical results for
$E_{{\cal N}}(\rho_{\rm out})$, whereas the dotted curves plot
$2\log_2(1+2^{E_{{\cal N}}(\rho_{\rm in}\rangle)/2})$. In the case of a pure
state these two quantities would be the same, according to
(\ref{LNa&bBS2p}), but for mixed states there is a small
difference, indicating the relation (\ref{LNa&bBS2p}) is a fair
approximation for mixed states.
\begin{figure}
\begin{center}
  \includegraphics[width=240pt,height=135pt]{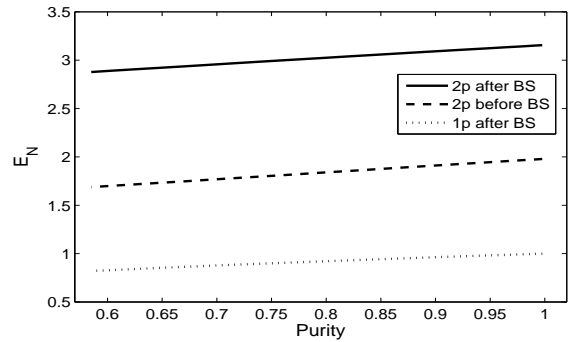}\\
  \caption{Logarithmic negativities of various mixed
  output states of a 50/50 beam splitter are plotted against purity.
  The single-photon result follows directly from
  Eq.~(\ref{LNm1}), while the two-photon case is
  calculated from state Eq.~(\ref{mix2pout}). We have chosen
  $\sigma=\Omega$ for both two-photon states here and in all remaining figures.
  }\label{LN-Pur}
\end{center}
\end{figure}

From the preceding section we know the precise relation between
entanglement generated by a beam splitter and purity  for
single-photon states. For comparison we plot in Fig.~\ref{LN-Pur}
the logarithmic negativity as a function of the purity of the
mixed state for both single- and two- photon states.  The almost
linear behavior indicates the close relation between the
characteristic purity of the system and the amount of entanglement
that can be extracted under ideal conditions. Obviously,
entanglement increases with purity.
\begin{figure}
\begin{center}
  \includegraphics[width=240pt,height=135pt]{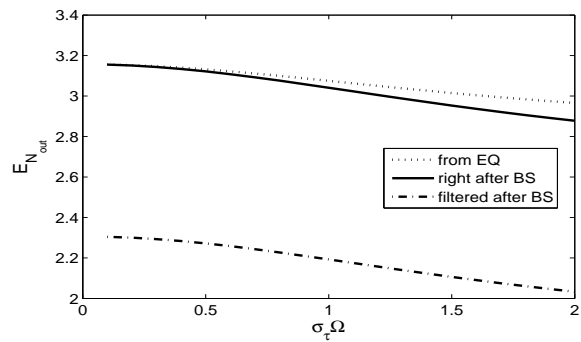}\\
  \caption{Logarithmic negativities for the output of  mixed
  two-photon states  (\ref{mix2pout})--(\ref{2pout}). The dotted
  line is calculated from the relation
  Eq.~(\ref{LNa&bBS2p}), which is true for a pure state only.
 The dash-dotted line is the entanglement that would result after
 local filtering of the output state (see Section \ref{filter}).
  }\label{LN_fil-sigma}
\end{center}
\end{figure}


\subsection{Adding the vacuum}
Type-II down conversion does not produce a two-photon state.
Instead it produces a superposition of a two-photon state and the
vacuum. Thus, let us first consider a pure state of the form
\begin{eqnarray}
  |{\rm vac2_{in}}\rangle=\sqrt{1-p}|{\rm v}\rangle
  +\sqrt{p}\sum_k\sqrt{\lambda_k}h_{k}^\dagger v_{k}^\dagger|{\rm
  v}\rangle,\label{vac_2p_in}
\end{eqnarray}
where  $p$ is the {\em a priori}  probability to detect two
photons, which typically will be small. In terms of the Schmidt
coefficients we can easily calculate the entanglement present in
the state (\ref{vac_2p_in}),
\begin{eqnarray}
  E(|{\rm vac2_{in}}\rangle)&=&-(1-p)\log_2(1-p)-p\log_2p\nonumber\\
  &&-p\sum_k\lambda_k\log_2\lambda_k,\label{vac_2p_Ein}\\
  E_{{\cal N}}(|{\rm vac2_{in}}\rangle)
  &=&2\log_2(\sqrt{1-p}+\sqrt{p}\sum_k\sqrt{\lambda_k}).
  \label{vac_2p_LNin}
\end{eqnarray}
Just as before we can express the entanglement of a state
resulting from splitting this state on a 50/50 beam splitter in
terms of the entanglement of the input state. The 50/50 beam
splitter accordingly converts the state into
\begin{eqnarray}
  |{\rm vac2_{out}}\rangle&=&\sqrt{1-p}|0\rangle|_c0\rangle_d+\frac{\sqrt{p}}{2}
  (|\tilde{2}\rangle_c|0\rangle_d+|0\rangle_c|\tilde{2}\rangle_d)\nonumber\\
  &&+\frac{\sqrt{p}}{2}\sum_k\sqrt{\lambda_k}(v_{ck}^\dagger h_{dk}^\dagger
  +h_{ck}^\dagger v_{dk}^\dagger)|{\rm
  v}\rangle,\label{vac_2p_out}
\end{eqnarray}
where $|\tilde{2}\rangle$ is given by Eq.~(\ref{tilde2_disc}). The
entanglement measures are calculated to be
\begin{eqnarray}
  &&E(|{\rm vac2_{out}}\rangle)\nonumber\\
  &=&-\frac{1-p/2+\sqrt{1-p}}{2}\log_2\frac{1-p/2+\sqrt{1-p}}{2}\nonumber\\
  &&-\frac{1-p/2-\sqrt{1-p}}{2}\log_2\frac{1-p/2-\sqrt{1-p}}{2}\nonumber\\
  &&+p-\frac{p}{2}\log_2p-\frac{p}{2}\sum_k\lambda_k\log_2\lambda_k,
  \label{vac_2p_Eout}
\end{eqnarray}
\begin{eqnarray}
  E_{{\cal N}}(|{\rm vac2_{out}}\rangle)=2\log_2(1+\sqrt{p}\sum_k\sqrt{\lambda_k}).
  \label{vac_2p_LNout}
\end{eqnarray}
We observe that
\begin{eqnarray}
  &&E_{\rm out}-\frac{1}{2}E_{\rm in}\nonumber\\
  &=&-\frac{1-p/2+\sqrt{1-p}}{2}\log_2\frac{1-p/2+\sqrt{1-p}}{2}\nonumber\\
  &&-\frac{1-p/2-\sqrt{1-p}}{2}\log_2\frac{1-p/2-\sqrt{1-p}}{2}\nonumber\\
  &&+\frac{1-p}{2}\log_2\left(\frac{1-p}{2}\right)+1,\label{vac_2p_Ediff}
\end{eqnarray}
and
\begin{eqnarray}
  2^{E_{{\cal N}{\rm out}}/2}-2^{E_{{\cal N}{\rm
  in}}/2}=1-\sqrt{1-p}.\label{vac_2p_LN diff}
\end{eqnarray}
These relations are the generalizations of Eq.~(\ref{Ea&bBS2p}) and
Eq.~(\ref{LNa&bBS2p}) respectively when vacuum is involved.

For mixed states involving the vacuum no analytical results seem
to be possible, so we reverted to numerical calculations.
Fig.~\ref{LN_fil-p_vac} plots some results from those calculations
as a function of the probability $p$.

\subsection{Local filtering}\label{filter}
The mixed two-photon state (\ref{mix2pout}) is written as a
mixture of pure two-photon states (\ref{2pout}). The latter states
are superpositions of two types of states: those with an odd
number of photons in each output port of the beam splitter
(namely, one), and those with an even number (namely, zero or
two). By imagining performing a quantum nondemolition measurement
of the parity of the photon number on (one of the) output ports,
we are applying a local filter. This filtering cannot increase the
entanglement and thus the entanglement after filtering gives a
lower bound to the total entanglement present in the mixed state
(\ref{mix2pout}).
\begin{figure}
\begin{center}
  \includegraphics[width=240pt,height=135pt]{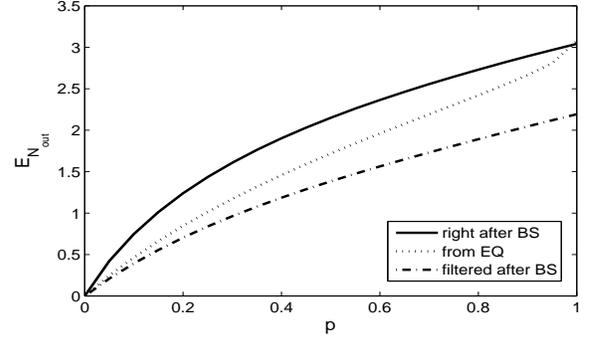}\\
  \caption{Logarithmic negativities for the output of a mixed
  two-photon state accompanied by vacuum at a fixed pump width
  $\sigma=\Omega$ and a mixture width $\sigma_\tau\Omega=1$. $p$ is the
  proportion of the two-photon state, as indicated in
  Eq.~(\ref{vac_2p_in}). The dotted line is calculated from the relation
  Eq.~(\ref{vac_2p_LNout}), which is true for a pure state only.
  The dash-dotted line is the entanglement after the beam splitter after
  the local filtering operation discussed in the text (Section \ref{filter}).
  }\label{LN_fil-p_vac}
\end{center}
\end{figure}
In the pure-state case we can calculate  the amount of
entanglement resulting from filtering by collapsing the output
Eq.~(\ref{vac_2p_out}) into either even-number-photon state
\begin{eqnarray}
  |\phi_{\rm even}\rangle=\frac{1}{\sqrt{1-p/2}}
  (\sqrt{1-p}|0\rangle|_c0\rangle_d\nonumber\\
  +\frac{\sqrt{p}}{2}|\tilde{2}\rangle_c|0\rangle_d
  +\frac{\sqrt{p}}{2}|0\rangle_c|\tilde{2}\rangle_d)
\end{eqnarray}
with probability $1-p/2$ or the odd-number-photon state
\begin{eqnarray}
  |\phi_{\rm odd}\rangle=\frac{1}{\sqrt{2}}
  \sum_k\sqrt{\lambda_k}(v_{ck}^\dagger h_{dk}^\dagger
  +h_{ck}^\dagger v_{dk}^\dagger)|{\rm  v}\rangle
\end{eqnarray}
with probability $p/2$. Averaging the entanglement over the two
possible measurement outcomes yields
\begin{eqnarray}
  &&E_{{\cal N}}^{\rm (even/odd)}\nonumber\\
  &=&(1-p/2)E_{{\cal N}}(|\phi_{\rm even}\rangle)+p/2E_{{\cal N}}(|\phi_{\rm odd}\rangle)\nonumber\\
  &=&p+p\log_2\sum_k\sqrt{\lambda_k}
  -\left(1-\frac{p}{2}\right)\log_2\left(1-\frac{p}{2}\right).
\end{eqnarray}
Figs.~\ref{LN_fil-sigma} and \ref{LN_fil-p_vac} illustrate this
filtering effect for {\em mixed} states as a function of
$\sigma_\tau\Omega$ for fixed $\sigma=\Omega$, and as a function
of $p$ for fixed $\sigma$ and $\sigma_\tau$, respectively. One
sees that about a third of the entanglement in the output state
arises from coherence between states like $|1,1\rangle$ and
$|0,2\rangle+|2,0\rangle$, in symbolic notation.

\section{Summary}\label{conclusions}
We have quantified the entanglement of various mixed states
containing exactly one or exactly two photons, as well as states
with a nonzero vacuum component, by using the logarithmic
negativity. For pure states we found simple relations between the
entanglement of single-photon and two-photon states before and
after a 50/50 beam splitter. For mixed states such relations are
still found to be approximately true.

The simplest result arises for a mixed delocalized single photon.
Its entanglement depends only on its purity. The result
illustrates that even a perfect deterministic single-photon source
(never producing more than a single photon) still may not be
sufficient for certain quantum-information processing purposes
(quantum computing based on dual-rail encoding, for example) if a
large degree of entanglement is needed.

We considered fairly realistic cases by explicitly including a
vacuum component of photon states, as well as including the
spectral and/or temporal shapes of photon wave packets. But an
obvious generalization of the present work would be to include
full three-dimensional mode structures. Moreover, in the case of
two photons impinging on a beam splitter we only treated the case
of photons with orthogonal polarizations (having in mind type-II
down conversion), but the similar case of identically-polarized
photons is interesting as well, and, perhaps surprisingly, more
complicated.

\section*{Appendix}
Here we calculate explicitly the logarithmic negativity for the first example to show the entanglement of a color-mixed polarization-entangled state is still one ebit, in spite of the mixed nature of the state. The two-photon singlet state, maximally entangled in polarization
--- horizontal(H) or vertical(V) ---, but equally and classically mixed
in color --- green(G) or blue(B)---, can be expressed in modes as
\[
\rho=\frac{1}{2}|\phi_1\rangle\langle\phi_1|+\frac{1}{2}|\phi_2\rangle\langle\phi_2|,
\]
where
\[
|\phi_1\rangle=\frac{1}{\sqrt{2}}(|GH\rangle_A|GV\rangle_B-|GV\rangle_A|GH\rangle_B),
\]
and
\[
|\phi_2\rangle=\frac{1}{\sqrt{2}}(|BH\rangle_A|BV\rangle_B-|BV\rangle_A|BH\rangle_B).
\]
Expanding $\rho$ in the basis of $(BH, BV)_A\otimes(BH,
BV)_B\oplus
(GH, GV)_A\otimes(GH, GV)_B$ gives
\[
\rho=\left(%
\begin{array}{cc}
  \begin{array}{cccc}
  0 & 0 & 0 & 0 \\
  0 & \frac{1}{4} & -\frac{1}{4} & 0 \\
  0 & -\frac{1}{4} & \frac{1}{4} & 0 \\
  0 & 0 & 0 & 0 \\
\end{array} & 0 \\
  0 & \begin{array}{cccc}
  0 & 0 & 0 & 0 \\
  0 & \frac{1}{4} & -\frac{1}{4} & 0 \\
  0 & -\frac{1}{4} & \frac{1}{4} & 0 \\
  0 & 0 & 0 & 0 \\
\end{array} \\
\end{array}%
\right).
\]
The partial transpose of $\rho$ in the same basis is then
\[
\rho^\Gamma=\left(%
\begin{array}{cc}
  \begin{array}{cccc}
  0 & 0 & 0 & -\frac{1}{4} \\
  0 & \frac{1}{4} & 0 & 0 \\
  0 & 0 & \frac{1}{4} & 0 \\
  -\frac{1}{4} & 0 & 0 & 0 \\
\end{array} & 0 \\
  0 & \begin{array}{cccc}
  0 & 0 & 0 & -\frac{1}{4} \\
  0 & \frac{1}{4} & 0 & 0 \\
  0 & 0 & \frac{1}{4} & 0 \\
  -\frac{1}{4} & 0 & 0 & 0 \\
\end{array} \\
\end{array}%
\right).
\]
The eight eigenvalues are easily found to be this: six times the eigenvalue $\frac{1}{4}$, and twice $-\frac{1}{4}$. This yields the logarithmic negativity of the state:
\[
E_{\cal N}=\log_2(6\times\frac{1}{4}+2\times|-\frac{1}{4}|)=1,
\]
as we announced.

\bibliography{yummyf}

\end{document}